# Of Substitution and Doping:
# Spatial and Electronic Structure in Fe Pnictides

Michael Merz[1]*, Peter Schweiss[1], Peter Nagel[1], Meng-Jie Huang[1], Robert Eder[1], Thomas Wolf[1], Hilbert von Löhneysen[1,2], and Stefan Schuppler[1]

[1]*Institut für Festkörperphysik, Karlsruhe Institute of Technology, 76021 Karlsruhe, Germany*
[2]*Physikalisches Institut, Karlsruhe Institute of Technology, 76131 Karlsruhe, Germany*

A highly intriguing aspect in iron-pnictide superconductors is the composition-dependent electronic structure, in particular the question if and how charge carriers are introduced to the system upon substitution of Ba by alkali metals or of Fe by other transition metals, *TM*. We report on a systematic study of spatial structure and electronic states by x-ray diffraction and x-ray absorption on a large number of compositions in the (Ba,K)(Fe,*TM*)$_2$As$_2$ family. The coherent combination of detailed structural information with an in-depth analysis of the electronic structure allows us to sensitively disentangle (charge-carrier) "doping" effects from "substitutional" effects. Results include a doping character that is site-decoupled, as well as *TM* 3*d* energy-level schemes that exhibit non-standard level sequences and even $t_2$-$e$ level crossings. Our study indicates that *doping per se* seems to play a lesser role than expected for pnictide superconductivity and magnetism.

## 1. Introduction

Recently, the layered iron-based pnictides and chalcogenides have emerged as an exciting new class of superconducting materials.[1] Most of these iron-based superconductors (FeSCs) share the same generic phase diagram reminiscent of other families of unconventional superconductors, with a dome-like shape of the SC phase boundary and a close proximity to magnetic phases. FeSC appears when the antiferromagnetic (AFM) and the structural phase transition temperatures ($T_N$, $T_s$) are significantly reduced – either by substitution on the pnictide (chalcogenide), the alkaline-earth, or the Fe site, or by pressure. Among the FeSCs, (Ba,K)(Fe,*TM*)$_2$(As,P)$_2$ ("122") is perhaps the best-studied pnictide system and may serve as a representative for all FeSCs. Although an understanding of phase diagram and physical properties of "122" and the other FeSCs has considerably advanced in the last years, essential issues for a detailed picture of these novel materials still remain unresolved.

In particular, it has been discussed controversially whether or not a *substitution* of Fe by

Mn, Co, Ni, or Cu does indeed lead to the widely assumed charge-carrier *doping* of (Ba,K)(Fe,*TM*)$_2$As$_2$: While some investigations[2–4] seem to be in favor of a picture where charge carriers arrive at the mostly Fe-derived bands near $E_F$, others[5–7] are not. The *TM* impurity may lead to a change of the Fermi surface, destabilizing magnetism in favor of superconductivity and also questioning a doping effect;[8] furthermore, if the total electron density peaks close to the impurity the latter *appears* isovalent to Fe.[8–11] For superconductivity and magnetism, observations such as Fermi-surface nesting, inter-band scattering between hole and electron pockets, and/or the appearance of a substitution-dependent nematic phase seem to play an essential role.[12] Further interesting observations include AFM order showing up for BaFe$_2$As$_2$ but disappearing or changing character upon Fe/*TM* or Ba/K substitution; or the fact that substitution of Fe by Mn or Cr, after initially suppressing AFM order, does not bring about superconductivity but leads to G-type AFM ordering; or the finding that BaFe$_2$As$_2$ and the Ba/K- and Fe/Mn-substituted systems show a more "bad-metal" behavior, unlike the other Fe/*TM* substitutions with their more "good-metal" character.

When trying to understand magnetism and superconductivity in FeSCs it thus appears necessary to separate direct doping effects from substitutional effects. We address this issue by combining single-crystal x-ray diffraction (XRD) and near-edge x-ray absorption fine structure (NEXAFS) on many compounds (from the same batches). The system under investigation is BaFe$_2$As$_2$, with partial substitution on the Ba site, on the Fe site, on both sites simultaneously ("bi-substituted"), or by two different *TM*s for Fe ("co-substituted").

2. **Experimental and Methods**

Single crystals for all compositions in this work were grown from self-flux in glassy carbon or Al$_2$O$_3$ crucibles as described elsewhere.[13] The resulting crystals typically have a surface with a size of about 5 × 5 mm$^2$ and a thickness around 1 mm. The composition of the samples was determined using energy-dispersive x-ray spectroscopy.

XRD data for more than 100 single crystals of various compositions were collected at room temperature on a Stoe four-circle diffractometer using Mo $K_\alpha$ radiation. All accessible symmetry-equivalent reflections were measured up to a maximum angle 2θ ≈ 95°. The data were corrected for Lorentz, polarization, extinction, and absorption effects. Using SHELXL,[14] around 300 averaged symmetry-independent reflections ($I > 4\sigma$) have been included for the respective refinements in space group *I*4/*mmm*. The refinement of all samples converged quite well and shows excellent weighted reliability factors ($wR_2$) which are typically around 4%.

NEXAFS measurements at the $L_{2,3}$ edges of Fe, Mn, Co, Ni, Cu and As were performed at the Institut für Festkörperphysik beamline WERA at the ANKA synchrotron light source (Karlsruhe, Germany). Using linearly polarized light, the spectra were taken simultaneously in bulk-sensitive fluorescence yield (FY) and in total electron yield (TEY) mode on single crystals that exhibit a mirror-like shiny surface. Photon-energy calibration to better than 30 meV was ensured by adjusting the Ni $L_3$ peak position measured on a NiO single crystal before and after each NEXAFS scan to the established peak position.[15] The spectral resolution was set to ≈0.3 eV for the Fe and TM $L_{2,3}$ edges and to ≈1.1 eV for the As $L_{2,3}$ edges.

To simulate the Fe and TM spectra, multiplet calculations were performed using the program CTM4XAS.[16] Spectra were calculated for different values of the crystal-field splitting and of the charge-transfer energy, taking the Hund's rule exchange interaction into account. Charge-transfer effects were included as well.

The Madelung potential or crystalline electric field, $V_{CEF}$, was calculated from a point-charge model by performing a multipole expansion. The $l=2$ and $l=4$ components determine the angular dependence of $V_{CEF}$ (and thus the energy-level scheme) and were obtained by real-space summation over spheres with a radius of 200 lattice constants, yielding good convergence. For these calculations, the point charges derived from the valences of our element-specific NEXAFS spectra and the bond distances from our XRD data were used. Ba (K) was assumed to have a valence of +2 (+1). The $l=2$ component turns out to be the dominant one in the layered structure of the pnictides because the contribution of a given ion at distance $r$ to the $l=2$ component is proportional to $r^{-3}$ whereas it is $r^{-5}$ for the $l=4$ component. To compute the energy-level scheme of an Fe/TM ion, values of $<r^2>$ and $<r^4>$ obtained from Hartree-Fock eigenfunctions for free Fe/TM ions with the appropriate valences were used.[17]

Relative level widths were roughly estimated from the approximate $r_p^{-4}$ and $r_d^{-5}$ dependence of the <Fe-As> and the <Fe-Fe> hybridization on the respective distances $r_p$ and $r_d$, see Ref. 18). In this approximation, possible differences between individual bands and their orbital origins are neglected.

## 3. Results and Discussion

First insight into substitutional effects in iron pnictides can be gained from the spatial structure. The "122" structure is shown in Fig. 1(a), and panels (d)–(f) plot the substitution-dependent evolution of key structural parameters – the pnictogen height above the Fe layer $\langle h_{As,Fe} \rangle$, the <Fe-As> bond length, and the <Fe-Fe> distance – versus a useful parameter $\Delta n$ that, for each substitution system, denotes the change in nominal electron count per formula unit relative to the parent compound.[19] For completeness, the lattice parameters and the <As-As> interlayer distance are included as well.

In the search for the underlying physics,[20,21] possible correlations between certain structure parameters and FeSC or its optimum $T_c$ have long been suggested. Often being discussed in this way is $\langle h_{As,Fe} \rangle$. From Fig. 1(d), we see that Ba/K substitution (gray circles) induces a linear increase of $\langle h_{As,Fe} \rangle$ (dashed line), and that *TM*-substituted, co-substituted, and bi-substituted systems with negative $\Delta n$ track it. Systems with positive $\Delta n$, however, deviate systematically and significantly from the dashed line. And superconductivity seems to turn up (filled symbols) or not (open symbols) quite independently from the absolute $\langle h_{As,Fe} \rangle$ value. A further suggested parameter is the averaged <Fe-Fe> distance, reflecting the in-plane $3d$ overlap between adjacent Fe/*TM* sites. As Fig. 1(e) shows, this parameter sometimes decreases (e.g. for Ba/K substitution), increases (Fe/Ni, Fe/Mn), or remains essentially constant (Fe/Co), and does not seem correlated with FeSC either. Neither does the <Fe-As> bond length, which reflects the hybridization strength between Fe/*TM* and As and is shown in Fig. 1(f). Finally, it has often been suggested that for FeSC to occur, the tetrahedron angle α should be close to its "ideal" value of 109.47°. And indeed, this ideal α is realized for (Ba,K)Fe$_2$As$_2$ with $\Delta n \approx -0.39$ where the highest $T_c^{max}$ (40 K) of the whole 122 family is found. For the sub-families with Fe/Co and Fe/Ni substitution, the respective $T_c^{max}$ (25 K, 18 K) appears for $\Delta n$ values (0.12, 0.18) where α, by the tetragonal distortion, is actually split into *two* α's: 111.7° and 108.4°. These observations seem to indicate a role of α for a "globally" *optimum $T_c$* for "122"; it appears less obvious whether the α splitting may put a boundary on the *range* where FeSC can occur. In any case, the comparisons between substitutional systems possible in this work seem to underline that the superconducting behavior observed is too complex and cannot adequately be traced back to a single structural parameter or to a combination of a small number of parameters.

For a closer look at the element-specific electronic structure, NEXAFS was performed at the Fe, *TM*, and As $L_{2,3}$ edges. At each edge, important information is present and can be extracted: energy shifts (for detecting small relative doping effects), multiplets (for fractional yet absolute valence states), and spectral weights (for charge-carrier transfer from/to the relevant element). [The Fe and Co spectra for Ba-122 are very similar to those for Sr-122 in Ref. 6), and are not shown.]

For iron, we focus on the *energy position* of the $L_3$-edge peak maximum. Fig 2(a) shows, as an inset, the enlarged Fe $L_3$ peak region for selected substitutions. Peak shifts $\Delta E$ derived from these and many more spectra are mapped versus $\Delta n$ and K substitution level in Fig.s 2(a) and (b), respectively. A clear trend of $\Delta E$ is visible only in Panel (b), indicating that substitution of Ba by K means (hole) doping at Fe 3$d$-derived states at $E_F$.[22] The plot $\Delta E$ vs. $\Delta n$, on the other hand [Panel (a)], shows that no matter what *TM* atom is substituted, co-substituted, or bi-substituted for Fe there is *no indication* for even a small *TM*-induced $\Delta E$: no doping to Fe 3$d$-derived states takes place.

For the *TM* substituents at the Fe site, Fig. 3 compares the *TM* $L_{2,3}$ NEXAFS to multiplet calculations,[23] giving a valence of ≈+1.0, ≈+2.3, and ≈+2.7 for *TM*=Mn, Co, and Ni, respectively. Cu resembles closely the 3$d^{10}$ configuration of Cu metal, compatible with a nominal valence of +1.

Remarkably, there is no "doping interference" in bi-substitution: even the substantial K content of the bi-substituted (Ba,K)(Fe,Mn)$_2$As$_2$ sample [blue symbols in Fig. 3(a)] does not affect the bulk Mn valence at all; the holes doped by K substitution for Ba all seem to go to Fe (see below).

For As, the spectral weight of the pre-edge region of As $L_3$ NEXAFS reflects the hole count in Fe/*TM*-As bonds with As 4$s$/4$p$ character,[24] and Fig. 4 plots this versus $\Delta n$. There is a systematic increase / decrease of spectral weight upon negative / positive $\Delta n$ for *TM* substitution, and no spectral-weight change for K substitution. In other words, K substitution leaves the electron count at As 4$s$/4$p$ states almost perfectly constant, whereas additional holes (electrons) are found at these states upon Fe/Mn (Fe/Co, Fe/Ni) substitution, respectively. The spectral-weight slope for Fe/Mn substitution is almost three times that for Fe/Co and Fe/Ni substitution.

The various doping effects found in Ba-122 are summarized in Table I. Most obvious is a strongly "disintegrated character" of doping and a surprisingly dominant role of As states in accommodating the doped charge carriers: Fe/Co and Fe/Ni substitution leads to electron doping solely in As 4$s$/4$p$ states and thus on As sites, while hole doping induced by Ba/K substitution occurs in Fe 3$d$ states and thus on Fe sites. And the Fe/Mn system, which dopes *holes* but exclusively to *arsenic* states, demonstrates that for Ba-122, not the *charge character* but the *substitution site* is decisive: Ba-site (Fe-site) substitution leads to Fe-site (As-site) doping. No "mixing" can be found; each type of doping leaves the "other" site unchanged – in all, a complete "site decoupling" of the electronic structure. A possibly related aspect was recently observed for (Ba,K)Mn$_2$As$_2$, where the magnetic behavior fully decouples between sites (localized AFM on Mn, itinerant FM on As).[25]

The level of electron doping, according to the fractional valences derived above from multiplets, is quite small: 0.6 per Co instead of the nominal value $\Delta n$=2, and 1.4 per Ni instead of 4; in each case only about $\Delta n/3$.[26] Cu, with its valence of ≈+1, does not even dope electrons at all. On the other hand, hole doping for Ba/K or for Fe/Mn substitution is efficient and transfers the complete $\Delta n$; this explains why the "Mn slope" in Fig. 4 is three times as steep as the "Co,Ni slope".

In addition, Table I also includes the Sr-122 systems studied previously.[6] Un-substituted SrFe$_2$As$_2$ exhibits slight but distinct differences to BaFe$_2$As$_2$ at both the Fe and the As $L_3$ edges: using Fig.s 2(b) and 4 for "calibration", Sr-122's Fe-$L_3$ $\Delta E$ of about 0.15 eV to lower energies indicates Fe-site electron doping by about 0.22 e, and its spectral-weight increase observed at As $L_3$ indicates As-site hole doping by about 0.28 h, compensating the Fe-site doping within the error limits. Compared to Ba-122 this may be a "self-doping" effect owing to the different equilibrium bond lengths and the reduced interlayer distance of ≈3.44 Å, and it sets a slightly changed "baseline" for substitution in the Sr-122 series. And indeed, Co substitution in Sr-122 is a little different: up to $\Delta n$ = 0.8, no doping effect was observed – neither at Fe nor at As –, indicating a Co valence in Sr-122 that is even more similar to that of Fe than in Ba-122.

We recall at this point that band structures for Ba-122, for instance in Ref. 9), show several Fe bands at $E_F$ and just a single (electron-like) band with significant As character,[27] leading to a relatively small As DOS at $E_F$. Within a rigid-band model, the element-specific doping would simply be distributed along the DOS fractions. How, then, may this "site decoupling" be understandable within a band-structure picture, for instance the observation that As-derived bands *dominate* doping for Fe/*TM* substitution? Important in this context might be that in such calculations, small changes in As position[28] can cause some hole-like bands to move significantly relative to the electron-like bands. The As height could then always adjust itself so that the combined filling of the Fe components in all electron and hole bands remains constant – favoring the As-dominated electron band that cannot be compensated in this way.

And the As height $<h_{As,Fe}>$ [Fig. 1(d)] does indeed change systematically with substitution. Of course, further aspects may be just as important – such as Hund's orbital correlations, or Lifshitz-like transitions with changing As height, affecting band character. [The s-shaped $<h_{As,Fe}>$ evolution for $\Delta n > 0$ may even be a consequence of the complicated band topology at $E_F$, see also Ref. 29).] More theoretical work on the As character of the near-$E_F$ bands is necessary, also based on DMFT[30] and taking structural as well as Hund's-metal aspects fully into account.

Moving on to a different aspect of the electronic structure, our results on spatial structure and valences enable us to calculate the relevant "energy-level scheme", the splitting of the Fe and *TM* 3*d* orbitals, from the full Madelung potential. Although a Madelung-potential calculation is pure electrostatics it does use the *complete* structure and turns out to capture essential physics: Fig. 5(a) shows for BaFe$_2$As$_2$, denoted as "Fe$^{2+}$", that the crystal-field terms arising from the layered character (more precisely: from breaking cubic symmetry) exceed by far the terms derived from the distorted tetrahedron alone, lifting the predominance of the $t_2$-$e$ splitting that has generally been assumed so far.[31] The level sequence for the tetragonal room-temperature (RT) structure starts with the "in-plane" $x^2$-$y^2$ and $xy$ orbitals, followed by the "out-of-plane" $3z^2$-$r^2$ and the degenerate $xz/yz$ states. The latter show a small splitting for the orthorhombic low-temperature (LT) structure.

Turning to the substitution-dependent evolution of the resulting energy-level schemes plotted in Fig. 5(b) we note first that for Co$^{2.3+}$, the level sequence is the only one *unaltered* from Fe$^{2+}$. Upon partial substitution, this allows CoAs$_4$ tetrahedra to be embedded well in the FeAs$_4$ matrix and indicates that the former, while being magnetic impurities, will show only little propensity (beyond some *k*-space broadening of the Co levels themselves) for acting as electronic impurities (scatterers). This appears consistent with the unchanged <Fe-Fe> distance in Fig. 1(e) and with the observation of a significant "coherent" peak in optics.[32]

For K$^+$/Fe$^{2.5+}$ the level sequence is again very similar; only the $3z^2$-$r^2$ states are now almost degenerate with the $xz/yz$ states. For the other substitutional systems, the in-plane orbitals move up in energy and eventually overtake the out-of-plane orbitals – an example of "level crossings". All these level shifts upon substitution appear far greater than previously discussed in literature. And the level crossings include some between $t_2$ and $e$ states, again highlighting the contrast to previous assumptions.

For partial Fe/Ni substitution, the orbital-level shift progresses much more quickly than for

Fe/Co. Both Ba(Fe$_{1-x}$Co$_x$)$_2$As$_2$ and Ba(Fe$_{1-x}$Ni$_x$)$_2$As$_2$ exhibit FeSC for a "$\Delta n$" approximately in the light-blue range of Fig. 5(b), and the scaling factor of ≈2 between their "superconducting domes" (plotted versus $x$) has long been interpreted as evidence for charge-carrier doping.[33] There may now be an alternative, scattering-based explanation: taking the increasing degree of "orbital and charge mismatch" when going from Fe$^{2+}$ via Co$^{2.3+}$ to Ni$^{2.7+}$ as a measure of a scattering cross-section, the total "scattering effect" of a certain amount of Ni substitution seems to be matched by the effect of roughly twice that amount of Co substitution.[34]

For Fe/Cu substitution, Cu appears essentially monovalent [*cf.* Fig. 3(c)], and with all Cu 3$d$ states being (nominally) occupied the As 4$s$/4$p$ states must play a significant role for hybridization and conductivity.[35] Furthermore, the 100% substituted "end member" [structure from Ref. 36)] suggests an extremely small As-As interlayer distance of 2.570 Å, much smaller than even for the "collapsed phase" in Ca-122, and a fairly 3D character.

For Fe/Mn substitution (leading to insulating BaMn$_2$As$_2$, "Mn$^+$"), everything points to a particularly localized character – the level crossing between the $x^2$-$y^2$ and $3z^2$-$r^2$ orbitals, the large <Fe-Fe> bond length, as well as the negligible hopping between Mn and As[37] – which may contribute to the metal-insulator transition and to the appearance of local magnetic moments (leading to G-type AFM order), unlike the itinerant magnetic character usually observed in the pnictides. Furthermore, bi-substituted Mn did not at all accept holes from K substitution, *cf*. Fig. 3(a), also consistent with Fig. 2 where these Ba/K-doped charge carriers are all found at Fe. All in all, Mn may be the prototypical example of a substituent that strives for (and achieves) localization in every possible way.

For the fully substituted systems, comparing the overall 3$d$-level *spreads* in Fig. 5(b) to the single-level *widths* may be able to describe some of the "bad-metal" *vs*. "good-metal" character observed, intertwined with "incoherent/coherent" metallicity and the degree of correlation effects. Estimated relative level widths are plotted in Fig. 5(c). For the Fe/*TM* substitution series, a systematic relationship is apparent, with Fe and Co exhibiting the largest widths and particularly small widths for Mn and Cu. Starting from BaFe$_2$As$_2$ which is generally considered a "bad metal",[32] BaMn$_2$As$_2$ and KFe$_2$As$_2$ then appear to be the two systems where the "bad-metal" character is plausibly even more pronounced: the former because the 3$d$ width decreases faster than the 3$d$ energy-level spread, the latter because the spread increases faster than the width. Slightly larger width plus somewhat smaller spread lends some "better-than-bad"-metal character to BaCo$_2$As$_2$; and for BaNi$_2$As$_2$ and BaCu$_2$As$_2$, the substantially reduced energy-level spread seems able to overcompensate a comparably modest decrease in estimated width, pointing to a fairly "good-metal" character that is indeed observed in these systems. All is consistent to transport measurements and to the "incoherent"–"coherent" character observed for these end members in optical reflectivity.[38]

## 4. Conclusion

Our systematic study of a great variety of substitution systems in the "122" family of FeSCs, combining XRD and NEXAFS for insight into their spatial and electronic structure, has shed new light on many important aspects: Fe/*TM* 3*d* energy-level schemes based on the full Madelung potential demonstrate that the basic level sequence is not the simple $t_2$-*e*-dominated one of a single Fe/*TM*-As tetrahedron but is considerably altered by the non-cubic, layered character of "122"; their evolution with substitution exhibits level crossings – even between $t_2$ and *e* states! – that may indicate "good" or "hampered" hybridization of the impurity tetrahedra (more or less propensity for impurity scattering); and even the SC domes for Fe/Co and Fe/Ni substitution can now be understood within a scattering picture. For the fully substituted compounds, 3*d* level spreads *vs*. level widths help to qualitatively describe their good-/bad-metal behavior, with everything that this implies. Charge-carrier doping is "site-decoupled": holes introduced by Ba/K substitution go only to Fe 3*d* states; carriers introduced by Fe/*TM* substitution go exclusively to As 4*s*/4*p* states. An intricate interplay between spatial structure and the character / origin of bands might be at work, with As 4*s*/4*p* bands playing a much more significant role for doping and charge transfer than previously suspected.

In all, this "site decoupling", the quasi-"reluctant" electron doping observed for Fe/Co and Fe/Ni, as well as the strong evolution of the energy-level scheme with substitution suggest that at least for Fe/*TM* substitution (and perhaps quite unexpectedly), charge-carrier doping ("doping *per se*") is less important for the electronic and magnetic structure of "122" pnictides than the scattering and structural effects of substitution.

**Acknowledgments**

We are grateful to R. Fromknecht, R. Heid, C. Meingast, B. Scheerer, and J. Schmalian for excellent support and fruitful discussions. We gratefully acknowledge the Synchrotron Light Source ANKA Karlsruhe and the Karlsruhe Nano Micro Facility (KNMF) for the provision of beamtime. Part of this work was supported by the German Science Foundation (DFG) in the

*E-mail: michael.merz@kit.edu

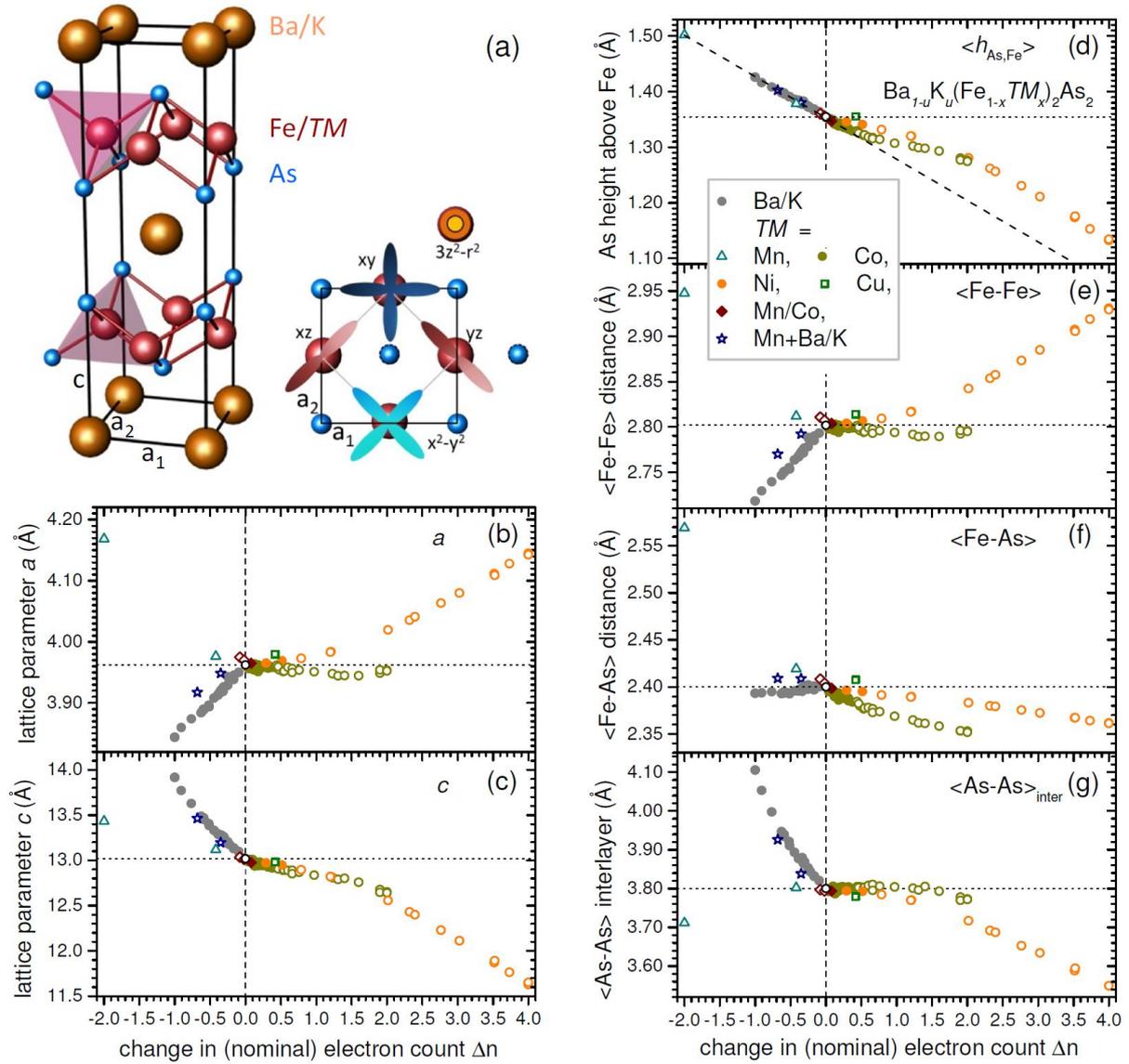

**Fig. 1** (Color) (a): Sketch of the unit cell of $BaFe_2As_2$ and substitution sites for room-temperature space group $I4/mmm$. Some of the edge-sharing Fe/*TM*-As tetrahedra are highlighted. Ba, Fe/*TM*, and As sit at the positions 0;0;0, 1/2;0;1/4, and 0;0;$z$, respectively. In the top view (to the right), the orbital orientation used is depicted together with the tetragonal axes. (b)–(g): Selected structural data of Ba-site and Fe-site substituted $BaFe_2As_2$. Samples which do (do not) exhibit superconductivity are represented by filled (open) symbols; $BaFe_2As_2$ is represented by an open black circle. The lattice parameters are depicted in (b) and (c). The substitution-dependent evolution of the As height above the Fe site, $<h_{As,Fe}>$, is displayed in (d), of the <Fe-Fe> distance in (e), and of the <Fe-As> bond length in (f). The <As-As> interlayer distance is shown in (g). Error bars (statistical) are of the order $10^{-4}$ Å, much smaller than the symbol size. The data are plotted as a function of the change in (nominal) electron count $\Delta n$ per formula unit for each substitution. All data shown are recorded at 300 K.

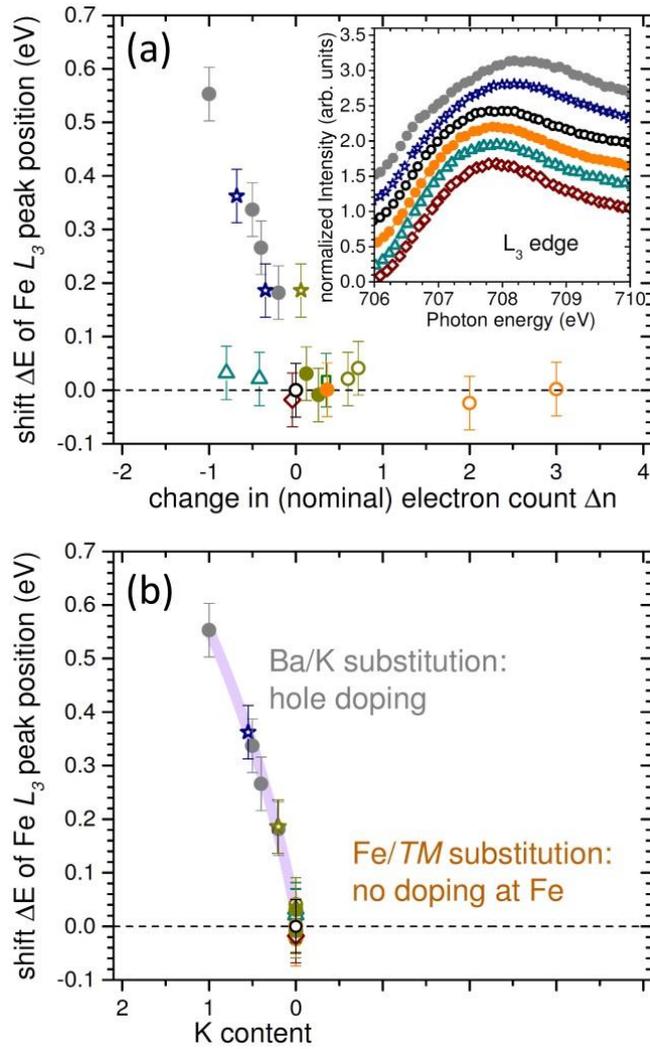

**Fig. 2**: (Color) Fe $L_3$ peak shifts $\Delta E$ (relative to the parent compound) mapped (a) versus $\Delta n$, and (b) versus K substitution level for the full set of substitutions. A clear and monotonic trend of $\Delta E$ is visible only in (b), indicating that substitution of Ba by K means (hole) doping at Fe $3d$-derived states at $E_F$. *TM* substitution, on the other hand, does not show any indication at all for a *TM*-induced $\Delta E$ and, thus, for doping at Fe sites. The inset in (a) depicts representatives for the normal-incidence FY Fe $L_3$ NEXAFS peaks. Symbols (colors, shapes) match those used for the structural data in Fig. 1. All data are recorded at 300 K. For clarity, the spectra are vertically offset.

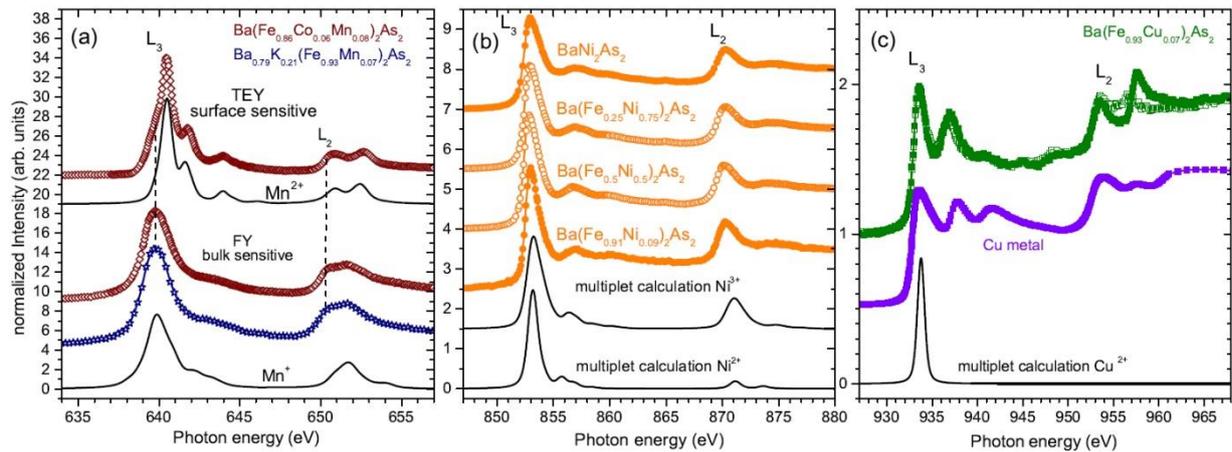

**Fig. 3** (Color) (a): TEY and FY Mn $L_{2,3}$ NEXAFS of co-substituted Ba(Fe,Mn,Co)$_2$As$_2$ and bi-substituted (Ba,K)(Fe,Mn)$_2$As$_2$. In the bulk of the sample, Mn is monovalent. Only in a thin surface layer does a Mn$^{2+}$ species dominate. K substitution does not seem to affect the Mn bulk valence, *cf.* the bi-substituted sample. (b) Ni $L_{2,3}$ NEXAFS of Ba(Fe$_{1-x}$Ni$_x$)$_2$As$_2$ for x = 0.09, 0.5, 0.75, and 1.0 detected in FY. The spectral shape is best described for a superposition of tetrahedrally coordinated Ni$^{3+}$ and Ni$^{2+}$ which leads to an effective valence of +2.7. The decreasing intensity with increasing Ni content is due to self-absorption and saturation effects. (c) Comparison of normal (closed symbols) and grazing (open symbols) incidence Cu $L_{2,3}$ NEXAFS of Ba(Fe$_{0.93}$Cu$_{0.07}$)$_2$As$_2$ recorded in FY. The spectral shape of the Cu $L_{2,3}$ edge fits reasonably well to Cu with a filled 3$d$ shell – Cu-metal spectrum – but definitely not to Cu$^{2+}$. Multiplet calculations in (a), (b), and (c) are plotted as lines. All spectra are recorded at normal incidence (unless indicated otherwise) and at 300 K. For clarity, the spectra are vertically offset.

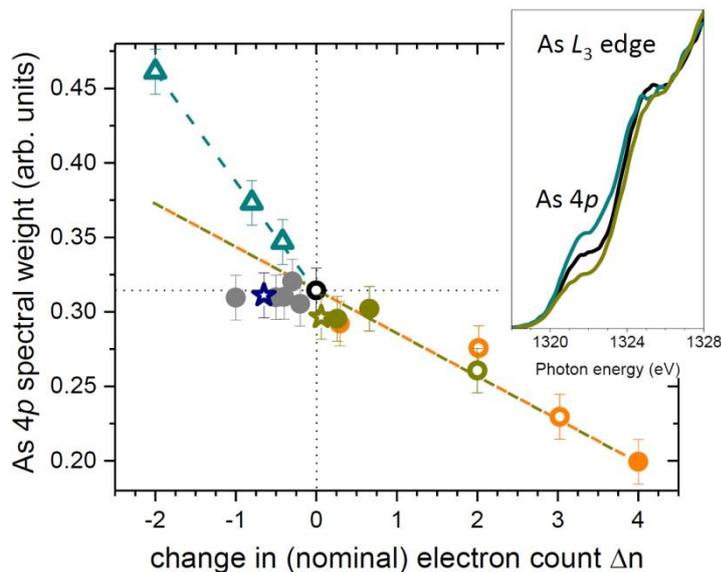

**Fig. 4**: (Color) Integrated spectral weight of the As 4$p$-derived pre-edge structure of the As $L_3$ absorption edge, see inset. For *TM*-site substitution, the spectral weight of the pre-edge increases upon Fe/Mn replacement (dark cyan triangles, $\Delta n < 0$) and decreases upon Fe/Ni and Fe/Co substitution ($\Delta n > 0$) while it remains almost unchanged by Ba/K substitution (grey circles, $\Delta n < 0$). Symbols (colors, shapes) match those used for the structural data in Fig. 1. The dashed lines are linear fits – one each for $\Delta n > 0$ and for $\Delta n < 0$ – through the spectral weights of the Fe-site substituted samples.

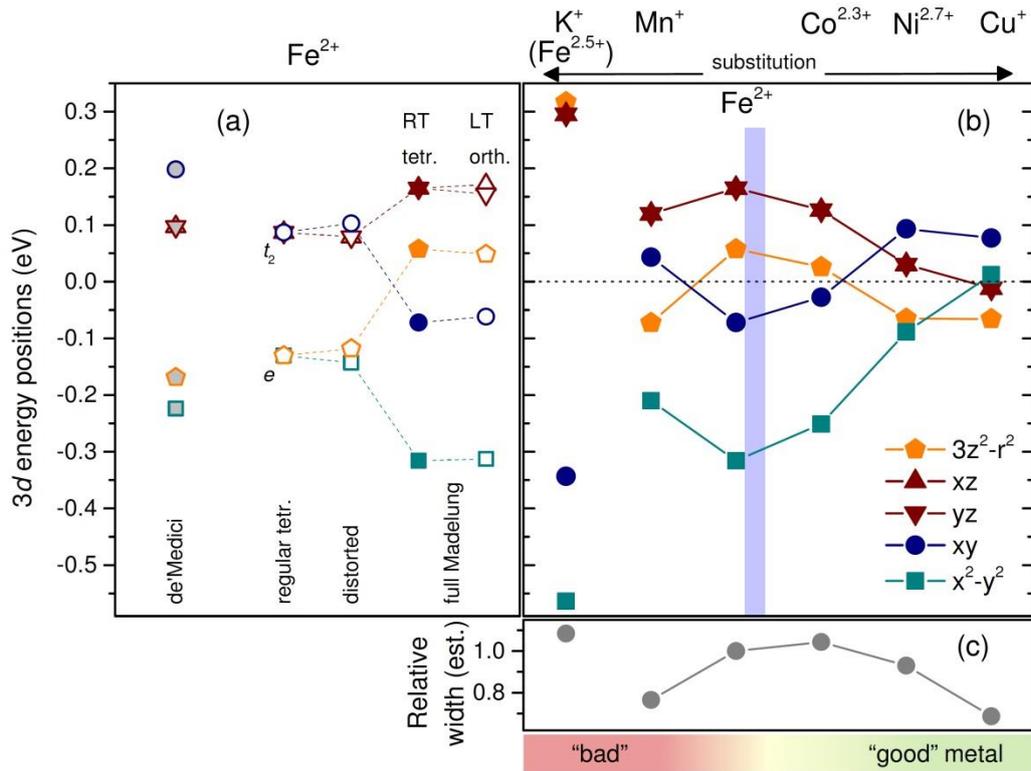

**Fig. 5** (Color) (a): Relative energies for the Fe 3$d$ orbital levels – "energy-level scheme" – for BaFe$_2$As$_2$ ("Fe$^{2+}$") and its evolution from a regular tetrahedron via a single distorted one to the full Madelung potential and finally to the orthorhombic splitting for the low-temperature structure. Juxtaposed at the far left is the energy-level scheme of Ref. 31) as a representative of the current state of the art. (b) Fe or *TM* 3$d$ energy-level schemes for Fe$^{2+}$ [as in (a)] and for the fully substituted compounds of this work. The Madelung potential and the *TM* valences derived from NEXAFS (this work) are taken into account. [For BaCu$_2$As$_2$ the structural data of Ref. 36) were used.] Relative to BaFe$_2$As$_2$ (Fe$^{2+}$), the sequence of the orbital levels remains unaltered for BaCo$_2$As$_2$ (Co$^{2.3+}$) and is only slightly modified for KFe$_2$As$_2$ (K$^+$/Fe$^{2.5+}$) while it is substantially changed for BaMn$_2$As$_2$ (Mn$^+$), BaNi$_2$As$_2$ (Ni$^{2.7+}$), and BaCu$_2$As$_2$ (Cu$^+$). Lines shown are guides to the eye. The approximate FeSC range for Fe/Co and Fe/Ni substitution is indicated in light blue. (c): Relative widths of the individual 3$d$ states estimated from the structural data (see text). Their relation to the energy spread of the orbital levels is reflected in the color bar at the bottom as the (expected and observed) "bad-metal" *vs.* "good-metal" behavior.

**Table I**: Overview of the most important doping effects observed in this work, relative to the un-substituted parent Ba-122; plus the Sr-122 results of Ref. 6). Doping effects of Sr-122 Co (Fe) are shown relative to Sr-122.

| System | Substitution | Doping Fe site | As site | Amount and comments |
|---|---|---|---|---|
| Ba-122 | Co (Fe) | – | e | $\approx 1/3\ \Delta n$ |
| Ba-122 | Ni (Fe) | – | e | $\approx 1/3\ \Delta n$ |
| Ba-122 | Mn (Fe) | – | h | $\approx \Delta n$ |
| Ba-122 | K (Ba) | h | – | $\approx \Delta n$ |
| Sr-122 | – | e | h | compensated "self-doping" by $\approx 0.25$ e and h, resp. |
| Sr-122 | Co (Fe) | – | – | $\Rightarrow$ valence Co $\approx$ valence Fe |